# Co-existence of Novel Ferromagnetic/Anti-ferromagnetic Phases in Two-dimensional Ti$_3$C$_2$ MXene

**Mehroz Iqbal,** [1, 2] **Jameela Fatheema,** [1] **Malika Rani,** [2] **Ren-Kui Zheng,** [3] **Saleem Ayaz Khan,** [4] **and Syed Rizwan**[1*]

[1]Physics Characterization and Simulations Lab, School of Natural Sciences (SNS), National University of Sciences and Technology (NUST), Islamabad 54000, Pakistan.
[2]Department of Physics, The Women University Multan, Multan 66000, Pakistan.
[3]State Key Laboratory of High Performance Ceramics and Superfine Microstructure, Shanghai Institute of Ceramics, Chinese Academy of Sciences, Shanghai 200050, China.
[4]New Technologies Research Centre, University of West Bohemia, Univerzitni 2732, 306 14 Pilsen, Czech Republic.

Corresponding authors: Syed Rizwan; syedrizwanh83@gmail.com
                      Saleem Ayaz Khan: sayaz_usb@yahoo.com

**Abstract**

This study reports first synthesis of MXene-derived co-existing phases. New family of two-dimensional materials such as Ti$_3$C$_2$ namely MXene, having transition metal forming hexagonal structure with carbon atoms have attracted tremendous interest now a days. We have reported structural, optical and magnetic properties of undoped and La-doped Ti$_3$C$_2$T$_x$ MXene synthesized using co-precipitation method. The c-lattice parameters (c-LP) calculated for La-MXene is c=18.3Å which is slightly different from the parent un-doped MXene (c=19.2Å), calculated from X-ray diffraction data. The doping of La$^{+3}$ ions shrinks Ti$_3$C$_2$T$_x$ layers perpendicular to the planes but expands slightly the in-plane lattice parameters. The band gap for MXene is calculated to be 1.06 eV which is increased to 1.44 eV after the doping of La$^{+3}$ ion that shows its good semiconducting nature. The experimental results for magnetic properties of both the samples have been presented and discussed, indicating the presence of ferromagnetic-antiferromagnetic phases co-existing. The results presented here are unique and first report on magnetic properties of two-dimensional carbides for magnetic data storage applications.

## 1. Introduction

Chemical bonding is a process to produce new compounds using periodic table elements through chemical reactions, forming new compositions and new structures. However, there are many phenomena observed when it comes to selective corrosion, dissolution, or oxidation conduction to configuring new structures and new phases of a compound. For example, Z. Zhang et al. reported the chemical and electrochemical de-alloying of metals resulting into nanoporous metals such as Ag‒Au, Cu‒Au, Au‒Zn, Pt‒Si or Cu‒Pt.[1] The de-alloying of aluminium (Al)-based alloys are more frequent candidates such as Al elimination from Cu‒Al‒Ag facilitating the formation of nonuniform passive films. Similarly, etching metals from metal carbides introduces new form of carbide derived carbon (CDC) with high specific surface area, having different dimensions and varying pore size. The selective metal extraction is simply based on the differences in reactivity of different elements during the compound formation under specific conditions and synthesis routes, which allows new materials with unique nanostructures.[2]

New two-dimensional (2D) materials namely MXene having transition metal carbides and carbonitrides, can be synthesized from their strongly bonded three dimensional (3D) layered parent compound by selective extraction of metal element. The 2D materials family include graphene, transition metal oxides, and transition metal carbides and nitrides (MXenes).[2] The 2D early transition-metal carbides and carbonitrides are synthesized by extracting the "A" layer sandwiched between the two "M" layers of transition metal in the MAX phases with general formula M$_{n+1}$AX$_n$ (n = 1, 2, or 3), where "M" is an early transition metal (Ti, Zr, Nb, Mo etc), "A" represents an A-group element

(mostly group IIIA and IVA), and "X" denotes C and/or N.[3] The MAX phases are a large (>60) family of layered, hexagonal ternary early transition metal carbides, nitrides, and carbonitrides.

The MAX powder is immersed in 39% concentrated hydrofluoric acid (HF) solution at room-temperature (RT) for different time durations for the selective extraction of "A" layer. The exfoliated product, called MXene, has large specific surface area with oxygen-containing functional groups (O and/or OH, and some fluorine) attached to its surface.[4] Dual nature from metallic to semiconductor and good mechanical to chemical stability shows hydrophilic nature which enhances the application importance of MXene in the area of photocatalysis, co-catalyst[5] sewag purification, polymer composites and energy storage devices, pseudocapacitors catalysis, hydrogen storage, and lithium-ion batteries.[6]

Chemical doping of heteroatoms is an efficient method to enhance the electrochemical, optical, electrical, magnetic, mechanical, thermal, photocatalytic, co-catalytic properties, and volumetric capacity of electrode materials. Notably, Y Tang et al, introduced the nitrogen-doped MXene to improve the electrochemical performance by increasing the conductivity of ions and creating surface-active sites, providing Faradaic reactions for additional collective capacity.[7] In addition, J. Luo et al. confirmed $Sn^{+4}$-doped MXene increases volumetric capacity and cyclic performance as anodes in Lithium-ion batteries.[7]

Alternatively, the room-temperature 2D materials with large surface area, good carrier concentration and ferromagnetic or antiferromagnetic properties are highly desirable. Since the last two decades, novel types of nanosheets-based nanomaterials have become an emerging field that provides new opening to understand this area especially for magnetic and spintronic devices. This is the subject of this study where we focus on the synthesis of novel 2D $Ti_3C_2T_x$ layered material to attain energetic and intrinsic room-temperature long-range ordered ferromagnetism with tunable semiconducting bandgap properties. This new direction is highly motivating into doping the $Ti_3C_2T_x$ sheets with different elements to study and enhance its ferromagnetic and semiconducting properties. Usually, rare-earth elements play major role in the development of different properties of the materials such as electronic, optical and magnetic enhancement. The present work is primarily based on doping rare-earth element namely Lanthanum La into $Ti_3C_2T_x$ sheets and to study its effects on the properties such as structural, electronic, optical and ferromagnetism.

## 2. Results and Discussion

### 2.1. Structural Characterization

The composition and overall crystallinity of MXene and La-doped MXene were determined by X-ray diffraction within the 2θ range from 5° to 80°. Overall, the characteristic peaks of $Ti_3C_2T_x$ were matched with previous reports[8], confirming successful synthesis of 2D material. The major peaks found in $Ti_3C_2T_x$ etched for 66 hours are at 2θ = 9.2°, 18.7°, 28.4°, 38.1°, 39.2, 40.2°, and 60.4° [JCPDS card:12-0539 ,9]. Within this list of peaks, structure of $Ti_3C_2T_x$ is represented at 9.2°, 18.7°, 38.1° and, 40.2° and the lattice parameters c and a are 19.2A° and 5.35A°, respectively. The lattice parameters were calculated using previous literature.[9-13] In addition, a small quantity of $Al_2O_3$ was observed at various peaks (34.5°, 38.1°, 39.2°, and 40.2°) that is attributed to the presence of $Al_2O_3$ nanocrystals on MXene surface which is not removed completely by etching.

The calculated values of lattice parameters "a" and "c" of MXene are 5.35Å and 19.2Å, respectively very close to values reported in literature.[14] The XRD pattern of La-doped MXene is shown in **Figure 1**. La is expected to form a complex compound on MXene surface via electrostatic interaction. The appearance of sharp peaks at 25.6°, 35.2°, 40.3°, 57.5°, 73.8° confirmed the presence of La on $Ti_3C_2T_x$ MXene surface. The ionic radius of $La^{+3}$ is 0.106 nm which is greater than the known 0.068nm and 0.058nm ionic radii of $Ti^{+4}$ and $Al^{+3}$, respectively. Hence, the placement of $La^{+3}$ in MXene structure is not easy to approach.[7, 15] An observable change in XRD shown in Figure 1 is obvious in the peaks of MXene 9.2°, 38.2° suppressed from its original sharpness and splits into two edges namely 9.2°, 9.3°, 38.2°and 38.4° shift from their original position at higher angle which is a signature that there is no or very little intercalation of $La^{+3}$. A peak at 60.7° shows a keen sharpness with a negligible back shift, investigation of which, results about basal plan, the stacks of MXene and increase in grain size.[16]

The lattice parameter (LP) calculated for La-doped $Ti_3C_2T_x$ shows values of about 5.36A° and 18.3A°, indicating a difference of 1.1A° in c-LP. MXene owing to –F, –O and –OH terminations would be the result of negatively charged surface which shows metallic nature of MXene[17]. This shifting could be originated from two possible effects: i) ion-exchange mechanism, ii) strong electrostatic interaction. So, the tendency of the ion exchange is greater at the surface of $[Ti-O]^-H^+$ due to the presence of –OH terminations, forming Ti–O–La or the negatively charged surface terminations –F, –O entraps $La^{+3}$ or with the exchange of weakly bonded $Al^{+3}$ present on the surface of MXene sheets. The partial ion exchange may also occur to some extent.[7, 18-19] The ion-exchange is observed due to weakly bonded $Al^{+3}$ ions. The prospect of decrease in c-LP is due to the strong electrostatic interaction between the abundantly negative charged surface and cationic $La^{+3}$ ions. Heavy positive ions are in access during the chemical reaction so they create a strong electrostatic potential around the MXene sheets that interacts with the negatively charged surface and causes the formation of complex electrostatic interaction between positive ions and negative terminations. This strong electrostatic interaction causes shifting of peaks at higher angle and make sure the presence of La$^{+3}$ because the radius of La$^{+3}$ is greater so the ability to lose electron is higher that creates more positive potential around the surface.[20] The crystalline sizes of the samples were calculated by well-known Scherrer's formula.

The mean crystalline size of pure MXene is about 33nm and that of La$^{+3}$-doped MXene is about 40nm. As a result, the doping of La$^{+3}$ enhance the crystal growth by replacing Al$^{+3}$ to some extent near the edges or surface of the sheet.[7] Table SI summarizes the XRD data.

## 2.2. Morphology, Surface and Elemental Analysis

Morphology, surface and elemental analysis evidences the constitution of nanoparticles and doping of La$^{+3}$ in MXene. Scanning electron microscopy is a technique introduced the morphology and phase analysis of a material e.g metals, semiconductors, nanoparticles etc. Investigation of 2D Ti$_3$C$_2$T$_x$ morphology confirmed that it is a layered structure in the form of stacks. These layered stacking is possible through the selective extraction of the Aluminium layer from its 3D parent compound 'MAX' (Ti$_3$AlC$_2$). The SEM images of Ti$_3$C$_2$T$_x$ MXene used here is shown elsewhere.[21] Small particle appeared on the surface of the MXene are usually reported as TiO$_2$ [22] but in this case, these surface particles are observed as Al$_2$O$_3$ phase [23] and is also confirmed by the XRD pattern. The SEM images **Figure 3a.** of La-MXene shows the loaded surface of MXene with Lanthanum oxide as well as represents the non-uniform stacking of the nanoparticles flakes. SEM micrography confirms the compact flakes like morphology.

To inquire the detail of structural changes Transmission Electron Microscopy TEM is used. **Figure 3a, b** shows the oxides loaded surface of La-MXene. The roughness of the surface increases with the embedded La$^{+3}$ ions at the surface of MXene (La-Ti$_3$C$_2$T$_x$). In Figure **3a** the roughness and pores of the La-doped MXene sheets are clearly observed. While to assess the structural changes of the La-doped MXene the specific area electron diffraction pattern (SAED) is observed in **Figure 3c** which confirms the hexagonal structure of nano thin lamellar of MXene Ti$_3$C$_2$T$_x$. MXene sheets retain their uniform hexagonal structure at most places in the High Resolution Transmission Electron Microscopy (HRTEM) but show some concentric rings due to the embedded La$^{+3}$ ions at the surface forming oxides as shown in **Figure 3c and 3d**. The La-doped NPs have an average granular size is 40nm with increased surface roughness.[40-443] The possible schematic diagram of the La-doped MXene is shown in Figure 4, representing that La$^{+3}$ ions are embedded at the surface adjusting with Ti$^{+4}$ particles.

**Figure 2** shows the EDX spectra of La-MXene indicating clearly the presence of La$^{+3}$ ions at the surface. The EDS results detect the elemental presence of Ti, Al, Si, C, O and F elements in which Ti and C are major elements. The amount of Al is very small which shows the definite exfoliation of MAX to give MXene and confirms the Al$_2$O$_3$ adhesive at the surface of Ti$_3$C$_2$T$_x$ sheets.[2,8,23-25] Doping of La$^{+3}$ in MXene is detected in this spectrum which shows the presence of La$^{+3}$ at more than three different energy levels. At 4.25 KeV La$^{+3}$ as oxide coexists with Ti$^{+4}$ at the surface and is strongly interacting which causes decrease in c-LP from 19.2 Aº to 18.3 Aº as was confirmed earlier also from the XRD pattern. Table SII summarizes the EDS data. It can be seen that the substitution of Al$^{+3}$ from the surface occurs because La$^{+3}$ content is increased compared to Al$^{+3}$, co-existing in the region of same energy thus, giving an indication of formation of La–O on MXene surface.

## 2.2. FTIR Spectra

Fourier transform infrared spectroscopy is a technique based on vibration of atoms, molecules or a compound. A comparison of FTIR results between MXene and La$^{+3}$-doped MXene confirms the doping of La$^{+3}$ and is shown in **Figure 5**.

The peak of FTIR spectra at 2161 cm$^{-1}$ confirms the formation of Ti$_3$C$_2$ [7] and peaks at 1977 cm$^{-1}$ and 2029 cm$^{-1}$ confirm the hexagonal structure of MXene. In the 2000–1700 cm$^{−1}$ range, a series of weak combination and overtone bands appear and the pattern of the overtone bands reflects the substitution pattern of the benzene ring. Skeletal vibration represented by C=C stretching absorbed in 1450-1600 cm$^{−1}$ range.[44-46] Some noise is observed in FTIR spectra from a range 1600-1750 cm$^{-1}$ and 3450-3850 cm$^{-1}$ which makes sure the presence of water molecules with stretching and bending of –OH group, respectively.[26] The deformation of Ti–O peak might occur at 550-750 cm$^{-1}$. After doping of La$^{+3}$ a sharp peak at 572 nm appears which confirm the presence of La–O and presence of heavy metal (La$^{+3}$) reflecting strong interaction with MXene at surface through positive electrostatic potential.[15]

## 2.3. UV-vis DRS Spectra

Diffuse reflectance spectroscopy was utilized to study the optical band gap shown in **Figure 6**. The UV-vis DRS of pure Ti$_3$C$_2$T$_x$ and La-doped Ti$_3$C$_2$T$_x$ is done in the range of 290-900 nm. The absorption spectra vs. wavelength of pure MXene and La-MXene exhibit specific regions. According to first-principles density function theory calculation, the inter- and intra-band transition arise Fermi energy levels for these absorption peaks.[6,27] The UV-vis spectra of La-MXene shown in **Figure 6a** demonstrates more absorption at longer wavelength near the red shift in the band gap transition compared with pure Ti$_3$C$_2$T$_x$ -MXene.

The absorption at longer wavelength 650nm-730nm, especially in the red spectra of visible spectrum can be characterized to the charge-transfer transition between rare earth ions 4f electrons and $Ti_3C_2T_x$ conduction and valence bands. As stated earlier, the La–O interacts with the surface of MXene which is the basis for more absorption in the red spectra because the semi core 4f levels are predominantly localized and do not take part in the chemical bonding and electronic conduction, but these levels and 4f electrons are easy to approach for more optical absorption and strong magnetic ordering.[2, 28] The light response range expands to the visible light, and the electron-hole pairs increased by the magnification of light absorption capability.[17, 29]

**Figure 6b** is an Arrhenius plot indicating that the absorption edge is caused by the direct permitted transition. The observed band gap for MXene and La-MXene is 1.19 eV and 1.44 eV, respectively. This demonstrates that the doping of La increases the amount of oxide on the surface of $Ti_3C_2T_x$ due to which, its band gap energy is increased and thus, absorbs large amount of energy in the visible spectrum. At larger band energy, terminations show more absorption in the visible region than for pristine MXene.[30] After La-doping, MXene behaves as a semiconductor because of the presence of oxides, O group differ from the other two functionalized groups as it has ability to gain two electrons from the structure to stabilize itself.[31, 2]

## 2.4. Co-existence of Magnetic Phases

MXene-$Ti_3C_2T_x$, having different terminations ($T_x$ = –O, –OH or –F), is prepared by etching $Ti_3AlC_2$-MAX at room temperature and then doped with Lanthanum ($La^{+3}$) to observe the change in its magnetic properties. **Figure 8** represents magnetization vs. temperature FC-ZFC curves for undoped and La-doped MXene. The results show that the FC-ZFC curve split under a low temperature range. It might be due to the oxidization of MXene surface because as Ingemar Persson et el. reported that –OH groups or O-1s at the surface alternates its adsorption sites due to the presence of influential –F and other impurities.[31] But as the temperature increases, it gains some stable sites and so the electronic cloud is shared more towards oxygen due to its more electronegativity than Titanium and Carbon. This may also be due to the presence of competing ferromagnetic/antiferromagnetic (AFM) phases. Shijun Zhao et al. reported computational results showing that the spatial spin density distribution of electrons with spin-up electrons are located at upper layer of Ti-3d while spin-down electrons forming electronic cloud on carbon between the layered structure of Titanium. The experimentally observed magnetic moment for $Ti_3C_2$ is 1.87 $\mu_B$.[33,34] Spin-up and Spin-down probably indicates the anti-ferromagnetic behavior of $Ti_3C_2$ while the spin-up behavior of Carbon and terminations ($T_x$ = –O, –OH and –F) attached with it exhibits an out-of-plan vibration which creates a weak ferromagnetic effect at the edges and surface.[35] Probably, the ferromagnetic domains can also be generated by defects or impurities like aluminium particles trapped in O-termination in the form of $Al_2O_3$.[36] From the magnetic behaviour of MXene, we infer that it is magnetically unstable at low temperature (5K, 10K) and is quasi-stable at higher temperatures (100K, 200K, and 300K). It was mentioned that there develops a considerable amount of electronic density of states near Fermi energy compared to the MAX phase due to redistribution of Ti-3d states from broken Al-Ti bonds.[34] This results in enhanced Ti-Ti bonding states near Fermi level. The Ti atoms on external sheet possess ferromagnetic ordering of spin-moments whereas they connect antiferromagnetically with opposite sheet containing titanium making it a mixture of FM-AFM phases. The internal Ti sheets remain non-magnetic making no contribution towards this anomalous magnetic origin.

**Figures 7 and 8** show magnetic hysteresis (M-H) curves for undoped and La-doped MXene at different temperatures. The undoped MXene shows a weak ferromagnetic (FM) response. Note that the M- H curves are unstable at 5K and is not shown here. The magnetic hysteresis loops for La-MXene, measured at 5K, 10K, 100K, 200K and 300 K, show high magnetic saturation and stable magnetic hysteresis loops. It can be seen that there exists asymmetry in M-H loops which is a clear indication of existence of exchange-bias effect. This means that the FM/AFM phases co-exist in the sample causing the exchange-bias effect to occur. This explains the splitting of FC-ZFC curves. After $La^{+3}$ substitution onto $Ti^{+4}$ site at the surface, it causes defects and disturbance in electronic distribution at the surface and also increase in the grain size which produces magnetization and increases the value of saturation magnetization ($M_s$) since the size of $La^{+3}$ is relatively larger than $Ti^{+4}$ cation. These results are clear indication of enhanced AFM behavior of doped-$Ti_3C_2T_x$ due to adsorption of $La^{+3}$ ions onto MXene surface. Thus, the material, due to co-existence of FM/AFM phases, can be termed to be multiferroic material for a wide range of temperature from 5K to 300K.

**Figure 8** summarizes the temperature dependence of exchange-bias (black) as well as coercivity (blue) calculated using the following formula[37]:

$$H_{eb} = (H_1 + H_2)/2, \quad \text{and} \quad H_{eb} = (H_1 - H_2)/2 \qquad (1)$$

Where, $H_1$ and $H_2$ are left and right coercive fields. It can be seen that the exchange-bias field is large at 10K but decreases with increase in temperature and remains non-zero at room-temperature. Similar trend can be seen for coercivity. This might be due to decreased contribution from FM phase at elevated temperatures due to spin-misalignment resulting in decrease in the FM-AFM interfacial area due to weakened interfacial interaction thus, reducing the loop-shift along one direction.[38]

## 2.5 Computational Analysis

For the analysis of La-doped MXene, the structure has also been studied using density functional theory (DFT). For DFT study, the calculations were performed using all electron full potential linear augmented plane wave (FLAPW) method in the framework of wien2k code [47]. The exchange correlation energy was solved using the Perdew, Burke and Ernzerhof generalized gradient approximation (PBE-GGA) functional[48]. The wave functions in the interstitial regions were expanded in plane waves, with the plane wave cutoff chosen so that $RMTK_{max}=8$ (where RMT represents the smallest atomic sphere radius and $K_{max}$ is the magnitude of the largest wave vector). The RMT radii were taken as 2.11 a.u. for Ti atoms, 1.73 a.u. for C atoms and 2.17 a.u. for La atoms. The wave-functions inside the spheres were expanded in spherical harmonics up to the maximum angular momentum $\ell_{max}=10$. The k-space integration was performed via a modified tetrahedron integration scheme. The self-consistencies of the ground state energies were obtained by using 10 k-points in irreducible Brillouin zone (IBZ).

The structure for the adsorption of La at a distance of 0.7 Aº from each other. Figures 11a and 11c show the FM structure with energy and figures 11b and 11d show the AFM structure with corresponding energies. From the energies of these four structures, we can say that the most stable structure are the structures with ferromagnetism for both situations, either placed at 0.7 Aº or 0.3 Aº. This is an indication that ferromagnetism is favoured over anti-ferromagnetism at 0K. Furthermore, the total density of states versus energy is given in figure 12. It is clear from figure that spin up density of states per eV is greater than spin down density of states per eV. Around the fermi energy level, the DOS is smaller as compared to the DOS at -4eV or 3eV but the behavior is still metallic and magnetic.

Table SIII shows the details of the magnetic moment of each atom in the supercell in terms of Bohr magneton. The total magnetic moment is 1.92 $\mu_b$ while at the interstitial position, the contribution is about 0.34672 $\mu_b$. La has a magnetic moment of ~0.05 while the magnetic moment of Ti is observed to vary according to its position in the cell such that Ti whose bond with Al had been broken results in higher magnetic moment. The magnetic moment of functional group elements is also presented, showing the presence of minimal amount of magnetism.

## 3. Conclusion

New family of two-dimensional materials called MXene with general formula $Ti_3C_2T_x$ was successfully fabricated after etching from its bulk powder namely MAX. We have reported experimental results on structural, optical and magnetic properties undoped and La-doped MXene. The lattice parameters changed after the doping indicating adsorption-dominant properties. The anomalous magnetization vs. temperature curves, ferromagnetic hysteresis loops and existence of exchange-bias indicate co-existence of ferromagnetic/antiferromagnetic phases at low and room-temperatures giving signature of multiferroic nature of our doped sample. The results presented here are novel and first report on magnetic properties of undoped $Ti_3C_2$ MXene and rare-earth Lanthanum doped MXene showing their potential for future two-dimensional carbide-based spintronic devices.

## 4. Materials and Methods

### 4.1. Materials

Hydrofluoric acid (HF, 39% concentrated), $Ti_3AlC_2$ (MAX) power, De-ionized water.

### 4.2. Synthesis of MXene and La-doped MXene

$Ti_3C_2$ was successfully prepared by selective etching Al layer from $Ti_3AlC_2$ using 39%-concentrated hydrofluoric acid (HF). Briefly, 10.0g of the as-prepared $Ti_3AlC_2$ MAX powder was immersed in 200ml HF solution under magnetic stirring at room-temperature for an optimized time of 66hrs. The resulting suspension was teemed and washed using deionized water until the pH value reached ~ 6. The suspension product was filtered using sterile membrane and washed with absolute ethanol. Finally, the resulting powder was dried in a convection oven at 50°C.[39]

The main process of synthesizing La-doped $Ti_3C_2T_x$ MXene is co-precipitation. The as-prepared $Ti_3C_2T_x$ as the raw material was added to different 250ml solutions of La-nitrate, and was stirred for 3-4hrs at 60-80°C. After that, the mixture was cold at room temperature and then filtered through sterile membrane. Final sample was washed for many times by using deionized water to maintain pH ~ 6. The as-synthesized powders were dried in oven at 50°C for further characterization.

### 4.3. Characterization

X-ray diffraction (XRD) was used to carry out structural analysis of prepared samples. The pattern was recorded by using Cu-Kα radiation of wavelength λ= 0.154 nm. Fourier-transform infrared spectroscopy (FTIR) was used to observe various bonds, scanning electron microscopy (SEM) was used to observe surface morphology and X-ray dispersive spectra to see elemental composition, ultra-violet (UV) spectroscopy was employed to measure optical bandgap and superconducting interference device (SQUID) was used to characterize magnetic properties of the samples.

## Supporting Information

Supporting Information is available on a separate sheet.

## Acknowledgement

The authors are thankful to Higher Education Commission (HEC) of Pakistan for providing research funding under the Project No.: 6040/Federal/NRPU/R&D/HEC/2016 and HEC/USAID for financial support under the Project No.: HEC/R&D/PAKUS/2017/783. The author also thanks School of Natural Sciences (SNS) at National University of Science & Technology The authors are thankful to Higher Education Commission (HEC) of Pakistan for providing research funding under the Project No.: (NUST), Islamabad, Pakistan for partial financial support. Saleem Ayaz Khan is grateful for support from CEDAMNF (CZ.02.1.01/0.0/0.0/15_003/0000358) of Czech ministry MSMT.

## Conflicts of interest

There are no conflicts to declare.

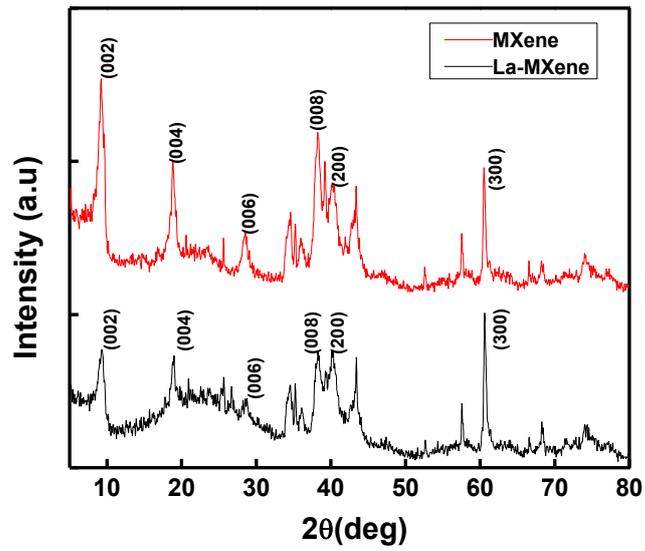

Figure 1: XRD patterns of MXene and La-doped MXene.

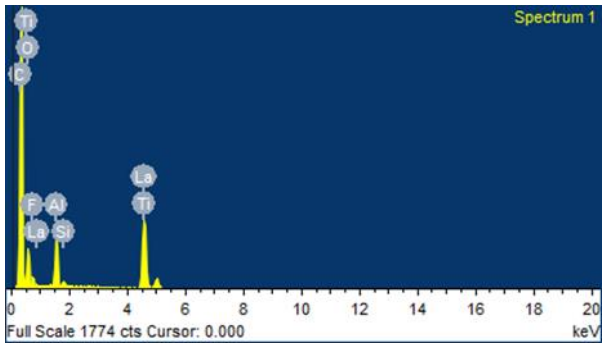

Figure 2: The EDS spectra showing elemental composition of La-doped MXene.

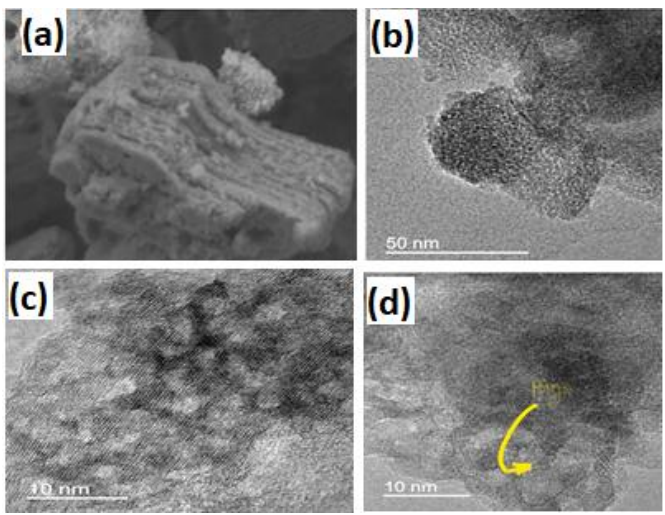

**Figure 3** (a) SEM images of La doped MXene sheets (b) TEM images of La doped MXene sheets (c-d) SAED image of La doped MXene Sheets

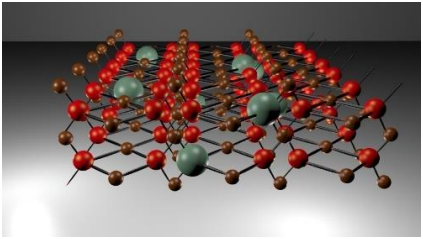

**Figure 4** Schematic representation of La embedded into the MXene sheets (grey=La, brown=C, red=Ti).

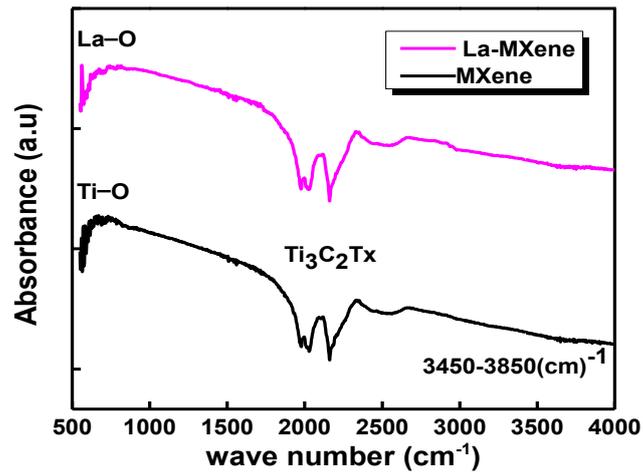

**Figure 5:** FT-IR spectra of samples before and after doping of La in MXene; Formation of La–O at the surface with adjustment of Ti–O at MXene surface is proposed.

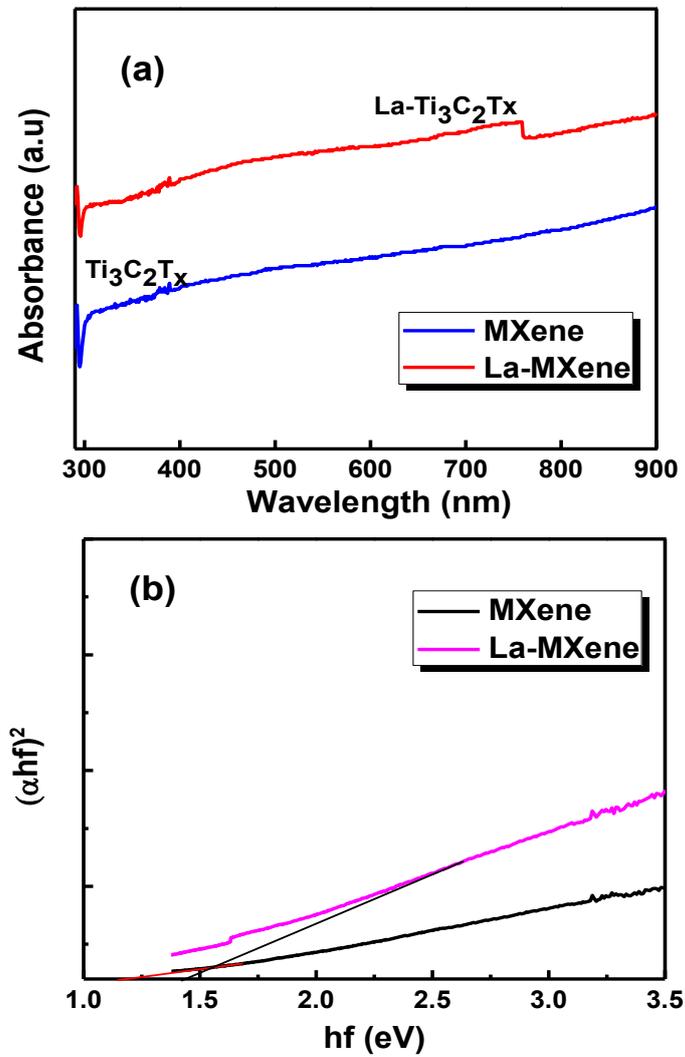

**Figure** 6. (a) UV-vis absorption spectra of $Ti_3C_2T_x$ and $La-Ti_3C_2T_x$. Absorption increases in red region after doping of $La^{+3}$ (b) band gap of $Ti_3C_2Tx$ and $La-Ti_3C_2T_x$.

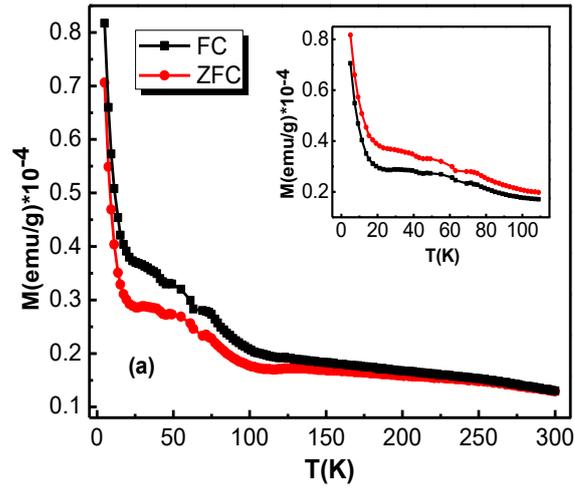

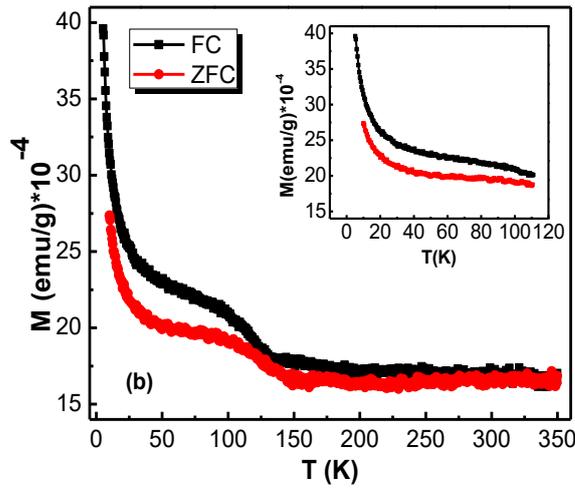

**Figure 7.** The FC-ZFC curves for (a) undoped MXene and (b) La-doped MXene.

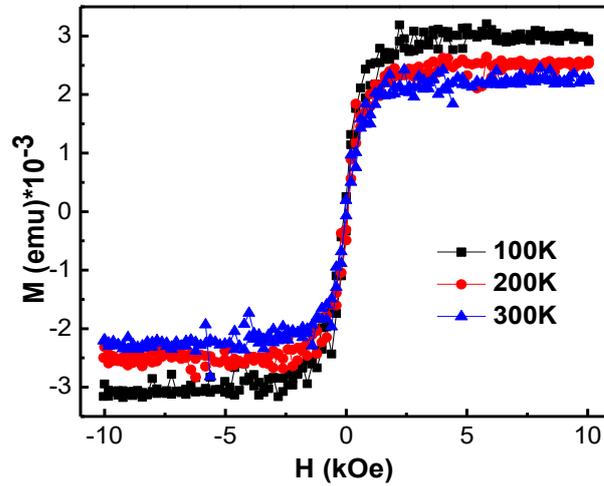

**Figure 8.** (a) M-H curves of undoped-MXene at 100K, 200K and 300K.

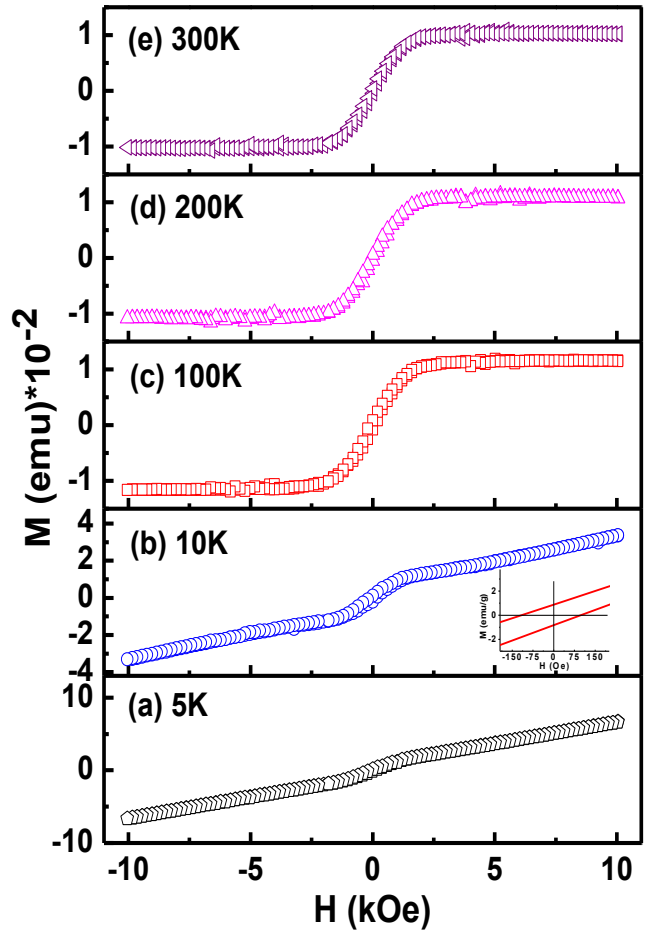

**Figure 9.** The M-H hysteresis loops for La-doped MXene measured at (a) 5K, (b) 10K, (c) 100K, (d) 200K and (e) 300K; Inset in (b) shows a clear exchange-bias effect.

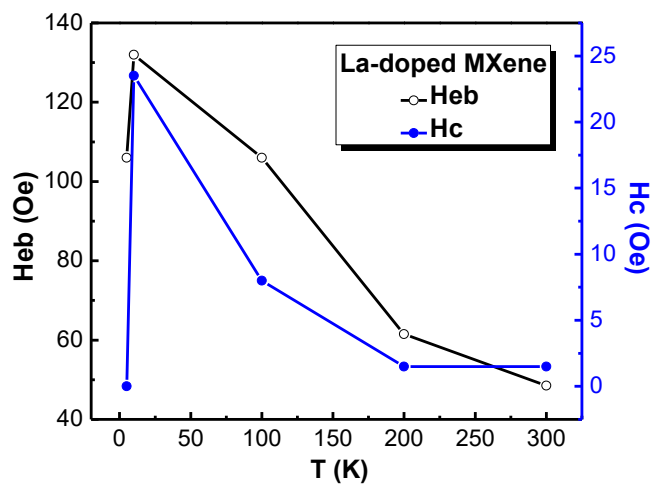

**Figure 10.** Temperature dependence of magnetic exchange-bias – Heb (black) and magnetic coercivity – Hc (blue) for La-doped MXene.

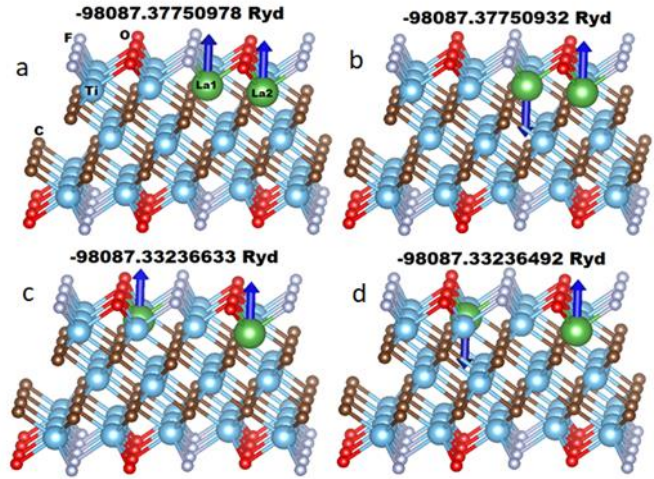

**Figure 11** structure of La doped Ti$_3$C$_2$-OF with different configurations and their corresponding ground state energies.

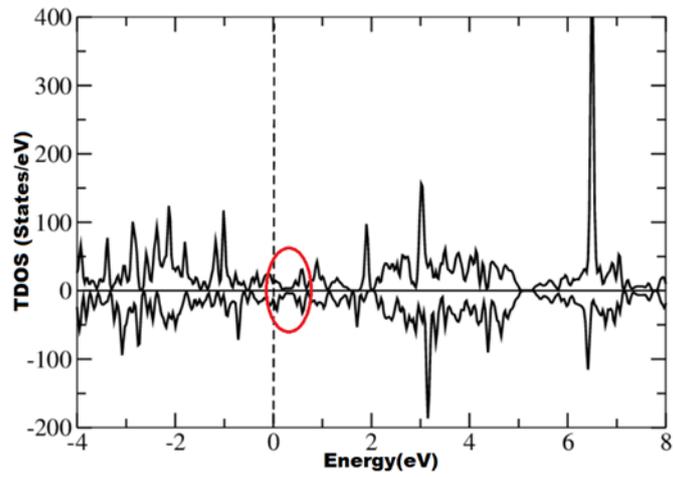

**Figure 12** total DOS versus energy of La doped Ti$_3$C$_2$-OF.